\documentstyle[11pt,newpasp,twoside,epsf]{article}
\def\deg{\hbox{$^\circ$}}
\def\arcmin{\hbox{$^\prime$}}
\def\arcsec{\hbox{$^{\prime\prime}$}}

\def\magsqarcsec{mag$/$\raisebox{-0.4ex}{\hbox{$\Box^{\prime\prime}$}\,}}
\def\sdeg{\raisebox{-0.4ex}{\hbox{$\Box^{\deg}$\,}}}

\def\cf{cf.~}

\def\eq{\!=\!}

\def\edcomment#1{\iffalse\marginpar{\raggedright\sl#1\/}\else\relax\fi}
\reversemarginpar

\begin{document}
\title{Tidal Streams around External Galaxies}
 \author{M.~Pohlen (1), D.~Mart{\'{\i}}nez-Delgado (2,1), 
S.~Majewski (3), C.~Palma (4), F.~Prada (1), \& M.~Balcells (1)}
 \affil{(1) Instituto de Astrof\'{\i}sica de Canarias, Spain  
         (2) Max-Planck-Institut f{\"u}r Astronomie, Germany
         (3) University of Virginia, USA 
         (4) Penn State University, USA}
\begin{abstract}
We have the unique opportunity to observe and model nearby streams 
around the two large Local Group spirals Milky Way and M31 in great 
detail. However, the detection of streams around other external galaxies 
is required to verify the general application of the derived results. 
We give a short summary of streams around other galaxies 
known in the literature, measuring for the first time 
the surface brightness of Malin's M\,83 stream with modern CCD imaging. 
In addition, we present four new detections of possible stellar streams 
around disk galaxies. 
\end{abstract}
\section{Introduction}
Hierarchical clustering scenarios of galaxy formation, such as cold dark
matter-dominated cosmologies, predict that structure 
forms first on small scales and later combines to form larger galaxies.
Low-mass objects similar to dwarf galaxies are supposed to be the building 
blocks and should have formed prior to the epoch of giant galaxy formation. 
Later they merge to build the larger galaxies. 
In this context, the accretion and tidal disruption of dwarf galaxies would 
play an important role in the evolution of galaxy halos and is essential 
for understanding the stellar mass assembly of large spiral galaxies.
Minor mergers (satellites with about 1/10 $M_{\rm parent}$) will 
significantly affect the star-formation and further evolution of 
disk galaxies, e.g.~by inducing disk thickening, warping, or bar formation.
Even smaller minor mergers, so called ``miniscule" mergers (Ibata 2001), 
with dwarf galaxies of lower mass should frequently happen to every 
large spiral galaxy. 
This accretion of low-mass systems, as a natural consequence of the 
merging history of a galaxy, is still ongoing and should give rise 
to long {\it stellar streams} (Johnston et al.~1996) around the 
parent galaxies.
\section{Known Stellar Streams and Open Questions}
The Local Group is the natural place to start 
looking for streams and the discovery of the prior unknown Sagittarius 
dwarf galaxy (Ibata et al.~1994) revealed a close satellite of the 
Milky~Way confirmed to be caught in the act of tidal destruction 
by means of star counts and kinematics.
In addition, Ibata et al.~(2001) detected another giant stellar stream 
within M31's (Andromeda Galaxy) halo. 
The Sgr and And stream support the scenario of ``spaghetti halos'' 
(Morrison et al.~2000): Galaxy halos build up by accumulation 
of tidally disrupted dwarf galaxies. 
But are tidal streams really common around galaxies?
Only a few detections of possible streams in external galaxies
have been made. Taking into account 
that Sgr viewed from outside would have a surface brightness of 
$\mu_{\rm V}\!=31$\magsqarcsec, this is not difficult to understand.
The first examples of extragalactic tidal streams are shown in the paper 
by Malin \& Hadly (1997). 
Using special contrast enhancement techniques they were able to 
highlight faint structures around nearby galaxies of large angular size 
on deep photographic plates.  From this paper M\,83
 (\cf Fig.~1) is probably one of the best cases, but structures 
in M\,104 (Sombrero) and NGC\,2855
may also represent stellar streams. 
\begin{figure}
\plotfiddle{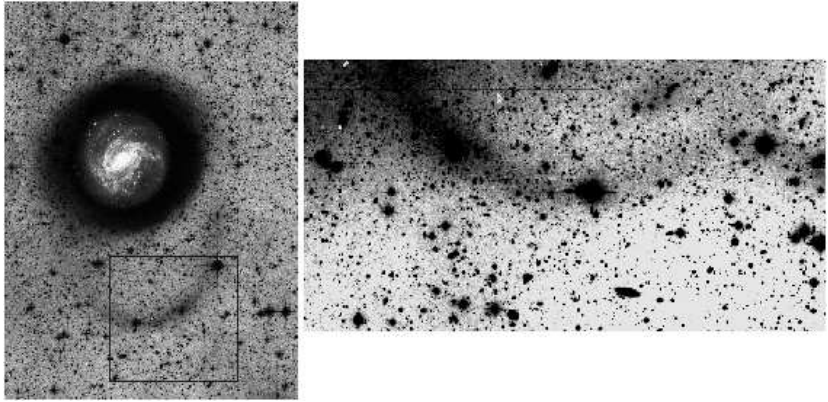}{4.5cm}{0}{38}{38}{-150}{-10}
\caption{M\,83: {\sl Left image} taken from Malin \& Hadley (1997). 
{\sl Right image} is a CCD exposure of the stream with the INT/WFC. The
box on the left marks approximately the position of the CCD chip.}
\end{figure}
Perhaps the most compelling example found to date is an elliptical 
loop, 45\,kpc in diameter, around the nearby, late-type, edge-on galaxy 
NGC\,5907 by Shang et al.~(1998).
Recent discoveries include Peng et al.'s~(2002) detection of a stellar 
stream in the envelope of the giant elliptical galaxy 
NGC\,5128 (CenA), and a fine example in the HST/ACS commissioning data
which nicely shows that UGC\,10214's long 
tail is produced by tidal disruption of an in-falling dwarf galaxy. 
The discovery of these extragalactic tidal streams evoke a number
of questions:   Are they exceptional cases or are tidal streams around 
galaxies really as common as predicted by current models?  Are these 
examples typical of the appearance of such streams?
What can we learn from these streams? What are the possibilities 
to find more of them? 
The latter question is addressed nicely by Johnston et al.~(2001), 
who estimate typical surface brightnesses of streams around external 
galaxies from semianalytic modelling. For one of their 
$10^8\,M_{\odot}$ model satellites the stream starts with surface 
brightness $\mu_{\rm R}\!\approx\!26.0$\,\magsqarcsec and, 
by spreading stripped material along its orbit, decreases
to 28.5\,\magsqarcsec after 1\,Gyr and down to 30.0\,\magsqarcsec 
after 4\,Gyr .
Systematic searches for more extragalactic tidal streams, 
in addition to providing evidence for the miniscule merger scenario, 
would hopefully turn up more examples of geometrically simple
cases like NGC\,5907 for which comparison to models like those of 
Johnston et al.~(2001) would yield reliable estimates of progenitor 
mass and age.
%
%
\section{New Streams and Outlook}
What methods can be used to detect extragalactic streams?  While previous 
discoveries have been serendipitous, the natural approach would be a deep 
survey of edge-on spiral galaxies in order to geometrically maximize the 
detection probability. This could be done either for nearby, large angular 
sized galaxies with the currently available wide-field imaging devices or 
for more distant samples using single CCD devices. Though the influence of 
the $(1\!+\!z)^4$ cosmological surface brightness dimming 
for galaxies out to 100 Mpc will be negligible --- only 0.1\,mag --- we 
have had little success with either approach: 
In a pilot survey using deep KPNO 0.9m MOSAIC images 
($\mu^{\rm R}_{\rm lim}\!\approx\!27$\,\magsqarcsec), we have not found
indications for streams around the nearby edge-on disks 
NGC\,3044, NGC\,3079, and NGC 3432, and, surprisingly, in a deep 
($\mu^{\rm V}_{\rm lim}\!\approx\!26-27$\,\magsqarcsec) 
imaging sample of $\approx\!80$ more distant, edge-on disk 
galaxies (Pohlen 2001) we have found only one clear example (ESO\,572-044) 
of a minor merger stream (two additional galaxies --- NGC\,5170 
and ESO\,286-018 --- show clear tidal features but are probably 
connected to more similar mass sized companions).
Meanwhile, we have had better success searching publicly 
available, deep, wide-field surveys. Though not typically centered on nearby
spiral galaxies and the typically detected streams will tend to be at 
redshifts where the surface brightness dimming becomes important  
(e.g., for a redshift of $z\eq0.12$ the brightness already decreases 
by 0.5\,mag), this technique has been used to find three of the four 
new stream candidates shown in Fig.~2: 
{\bf ESO\,572-044}:
This Sb galaxy (diameter: $d\eq$1.7\arcmin) is probably surrounded 
by a tidal stream. It is enclosed by an arc-like structure with a 
possible dwarf galaxy as the origin. ESO\,572-044 exhibits a large 
thick outer disk at low surface brightness which may indicate a 
recent interaction with the similar-sized nearby spiral ESO\,572-046 
($\Delta v_{\rm rad}=150$kms$^{-1}$). 
{\bf PGC\,1018343}:
This small galaxy ($d\approx\!25$\arcsec) was fortuitously found  
on a deep ING/WFC image. It is listed as NPM1G-07.0455 in NED and 
also detected in 2MASS, but has no measured redshift. The stream 
is clearly visible on the lower right side but also extends to 
the upper left side. 
{\bf PGC\,751050}:
This edge-on S0 galaxy ($d\approx\!25$\arcsec) is another serendipitous 
detection of a possible stream found in the ESO-WFI wide-field image of 
the Chandra Deep Field South. The stream is obvious and connected to a 
possible satellite at the end. It has a 2dF redshift 
of $z\eq0.1028$ and is listed in 2MASS as 2MASXi J0332467-274212. 
{\bf S\,J160610.7+552700 }:
This spiral galaxy ($d\approx\!20$\arcsec) with unknown distance 
lay by chance in the field of the HST image of the famous Tadpole galaxy 
taken during ACS commissioning. It also shows a stream extending on both 
sides. 
We are in the process of collecting the missing redshifts and 
calibrated photometry that will allow a detailed comparison of these new 
detections with the Johnston et al.~(2001) models. 
In addition, we are working with a deep, partly multi-colour, public survey 
of 10\,\sdeg from the Wide Field Camera at the ESO 2.2m telescope 
(kindly provided by M.Schirmer and T.Erben from Bonn University). This 
survey covers a set of random fields outside of any specific cluster, 
and therefore provides a suitable sample of isolated spiral galaxies at 
various redshifts to systematically search for tidal streams. 
%

%
{\bf Acknowledgments.  }
Credits for UGC\,10214 to NASA, H. Ford (JHU), G. Illingworth (USCS/LO), 
M.Clampin (STScI), G. Hartig (STScI), the ACS Science Team, and ESA. 
We would like to thank Mischa Schirmer and 
Thomas Erben at the ``Wide Field Expertise Center" of the 
Institut f\"ur Astrophysik und Extraterrestrische Forschung der Universit\"at 
Bonn (IAEF) in Germany for providing the reduced ESO/WFI CDF-S image. 
In addition, we would like to thank D.~Malin for 
providing the deep photographic image of M83 (Fig.~1).
\begin{figure}
\plotfiddle{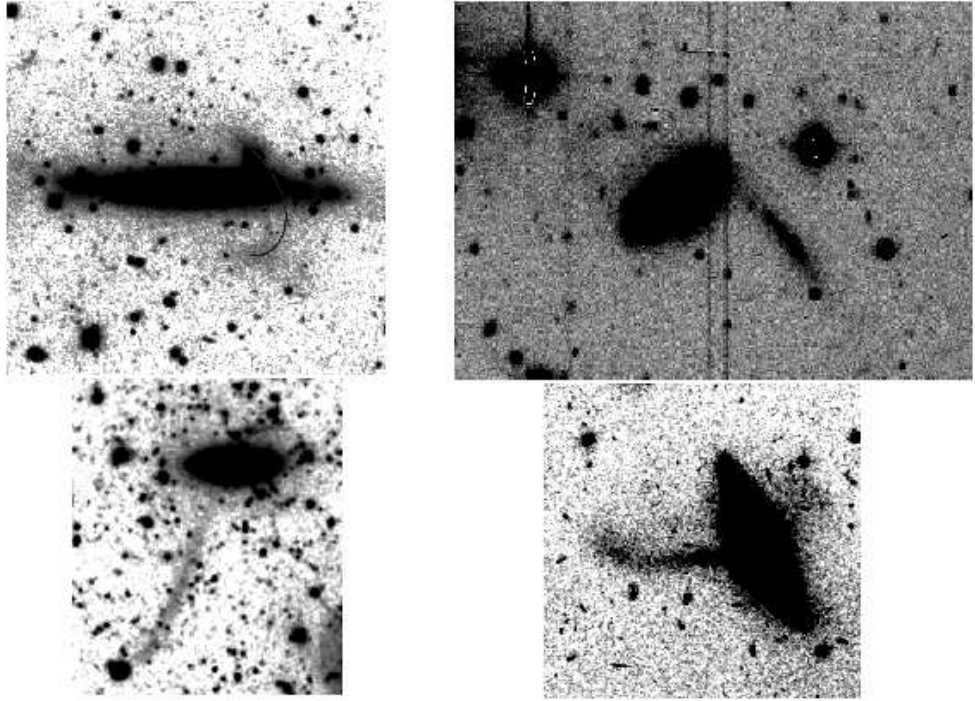}{8.6cm}{0}{45}{45}{-160}{-10}
\caption{New streams:
ESO\,572-044, ESO\,1.5D/\-DFOSC, 60\,min, $V$-band image 
{\sl (upper left panel)}; 
PGC\,1018343, INT/WFC, 30\,min, $r$ {\sl (upper right)}; 
PGC\,751050, ESO/WFI, 17.8\,hour, $R$ {\sl (lower left)};
S\,J160610.7+552700, HST/ACS, 134\,min, $F606W$ {\sl (lower right)} 
}
\end{figure}
%


\begin{references}
\reference Ibata, R.~A., Gilmore, G., \& Irwin, M.~J.\ 1994, Nature, 370, 194 
\reference Ibata, R.\ 2002, ASP Conf.~Ser.~275: Disks of Galaxies, p.431 
\reference Ibata, R., Irwin, M., Lewis, G., Ferguson, A.~M.~N., \& Tanvir, 
N.\ 2001, Nature, 412, 49 
\reference Johnston, K.~V., Hernquist, L., \& Bolte, M.\ 1996\ ApJ, 465, 278
\reference Johnston, K.~V., Sackett, P.~D., \& Bullock, J.~S.\ 2001, ApJ, 
557, 137
\reference Malin, D.~\& Hadley, B.\ 1997, PASA, 14, 52
\reference Morrison, H.~L., Mateo, M., Olszewski, E.~W., et al.~2000, 
AJ, 119, 2254
\reference Peng, E.~W., Ford, H.~C., Freeman, K.~C., \& White, R.~L.\ 2002, AJ, 124, 3144 
\reference Pohlen, M.\ 2001, PhD thesis, University Bochum
\reference Shang, Z., et al.~1998, ApJL, 504, L23; 
\end{references}
\end{document}